\documentclass[preprint]{elsarticle}

\usepackage{lineno}
\modulolinenumbers[1]
\usepackage{hyperref} 

\usepackage{soul}

\usepackage{lipsum}
\usepackage{rotating}
\graphicspath{{figures/}}
\usepackage{pstricks, pst-node, psfrag}
\usepackage{amssymb,amsmath}
\usepackage{mathtools}
\usepackage{verbatim,enumerate}
\usepackage{rotating, lscape}
\usepackage{setspace}
\usepackage[hang, flushmargin]{footmisc}
\usepackage{subfig}
\usepackage{caption}
\usepackage{cancel}
\usepackage{float}
\usepackage{ragged2e}
\usepackage{centernot}
\usepackage{tikz}
\usepackage{tikzsymbols}
\usetikzlibrary{shapes,arrows,positioning,plotmarks}
\usepackage{textcomp} 
\usepackage{gensymb} 

\tikzstyle{endpt} = [rectangle, draw, fill=red!20,
    text width=9.8em, text centered, rounded corners, minimum height=4em]
\tikzstyle{block} = [rectangle, draw, top color=white, bottom color=blue!20,
    text width=9.8em, text centered, rounded corners, minimum height=4em]
\tikzstyle{line} = [draw, -latex', very thick]

\usepackage{listings}
\usepackage{color} 
\definecolor{mygreen}{rgb}{28,172,0} 
\definecolor{mylilas}{rgb}{170,55,241}
\definecolor{mygray}{rgb}{0.5,0.5,0.5}
\definecolor{mycyan}{rgb}{0,255,255}
\definecolor{magenta}{rgb}{1,0,1}

\definecolor{backgreen}{rgb}{0.00, 0.169, 0.212}
\definecolor{textgray}{rgb}{0.514, 0.580, 0.589}

\usepackage[T1]{fontenc}
\newcommand*\patchAmsMathEnvironmentForLineno[1]{%
    \expandafter\let\csname old#1\expandafter\endcsname\csname #1\endcsname
    \expandafter\let\csname oldend#1\expandafter\endcsname\csname end#1\endcsname
    \renewenvironment{#1}%
        {\linenomath\csname old#1\endcsname}%
        {\csname oldend#1\endcsname\endlinenomath}}%
\newcommand*\patchBothAmsMathEnvironmentsForLineno[1]{%
    \patchAmsMathEnvironmentForLineno{#1}%
    \patchAmsMathEnvironmentForLineno{#1*}}%
\AtBeginDocument{%
    \patchBothAmsMathEnvironmentsForLineno{equation}%
    \patchBothAmsMathEnvironmentsForLineno{align}%
    \patchBothAmsMathEnvironmentsForLineno{flalign}%
    \patchBothAmsMathEnvironmentsForLineno{alignat}%
    \patchBothAmsMathEnvironmentsForLineno{gather}%
    \patchBothAmsMathEnvironmentsForLineno{multline}%
}

\newcommand{\R}{\mathbb{R}}

\newcommand{\cO}{\mathcal{O}}

\newcommand{\cU}{\mathcal{U}}

\newcommand{\defeq}{\coloneqq}

\newcommand{\pder}[2][]{\frac{\partial#1}{\partial#2}}

\newcommand{\dt}{\Delta t}
\newcommand{\dx}{\Delta x}

\journal{Advances in Water Resources}

\biboptions{sort&compress}
\begin{document}

\begin{frontmatter}

\title{Reactive Particle-tracking Solutions to a Benchmark Problem on Heavy Metal Cycling in Lake Sediments\tnoteref{mytitlenote}}
\tnotetext[mytitlenote]{This work was partially supported by the US Army Research Office under Contract/Grant number W911NF-18-1-0338; the National Science Foundation under awards EAR-1417145 and DMS-1614586; and the DOE Office of Science under award DE-SC0019123.}

\author{Michael J. Schmidt\fnref{nd}}
\ead{mschmidt1@mines.edu}
\author{Stephen D. Pankavich\fnref{ams}}
\ead{pankavic@mines.edu}
\author{Alexis Navarre-Sitchler\fnref{hydro}}
\ead{asitchle@mines.edu}
\author{Nicholas B. Engdahl\fnref{wsu}}
\ead{nick.engdahl@wsu.edu}
\author{Diogo Bolster\fnref{nd}}
\ead{bolster@nd.edu}
\author{David A. Benson\fnref{hydro}}
\ead{dbenson@mines.edu}

\fntext[nd]{Department of Civil and Environmental Engineering and Earth Sciences, University of Notre Dame, Notre Dame, IN, 46556, USA}
\fntext[ams]{Department of Applied Mathematics and Statistics, Colorado School of Mines, Golden, CO, 80401, USA}
\fntext[hydro]{Hydrologic Science and Engineering Program, Department of Geology and Geological Engineering, Colorado School of Mines, Golden, CO, 80401, USA}
\fntext[wsu]{Department of Civil and Environmental Engineering, Washington State University, Pullman, WA, 99164, USA}

\begin{abstract}

Geochemical systems are known to exhibit highly variable spatiotemporal behavior.
This may be observed both in non-smooth concentration curves in space for a single sampling time and also in variability between samples taken from the same location at different times.
However, most models that are designed to simulate these systems provide only single-solution smooth curves and fail to capture the noise and variability seen in the data.
We apply a recently developed reactive particle-tracking method to a system that displays highly-complex geochemical behavior.
When the method is made to most closely resemble a corresponding Eulerian method, in its unperturbed form, we see near-exact match between solutions of the two models.
More importantly, we consider two approaches for perturbing the model and find that the spatially-perturbed condition is able to capture a greater degree of the variability present in the data.
This method of perturbation is a task to which particle methods are uniquely suited and Eulerian models are not well-suited.
Additionally, because of the nature of the algorithm, noisy spatial gradients can be highly resolved by a large number of mobile particles, and this incurs negligible computational cost, as compared to expensive chemistry calculations.

\end{abstract}


\end{frontmatter}


\section{Introduction} 
\label{sec:introduction_hMetal}

Chemical reactions are ubiquitous in hydrologic systems and play a controlling role in the small and large scale behavior of many systems of practical interest \cite[e.g.][]{dentz2011mixing,valocchi2018}.
However, predicting complex reactions in realistic environmental settings, which are typically characterized by high degrees of heterogeneity and uncertainty with multiple processes occurring at different spatial and temporal scales, still remains a demanding challenge.
Many theoretical \cite[e.g.][]{LeBorgne2010,kitanidis_dilution,Engdahl_PRE,Barros2012} and computational approaches \cite[e.g.][]{steefel_RTM,mayer_multiCompRTM,Beisman2015,Benson_react} have and continue to be developed to tackle this issue.
On the computational side, methods for simulating reactive transport broadly fall into two categories: Eulerian and Lagrangian.
Defined in broad terms, Eulerian methods are grid-based methods, while Lagrangian methods are gridless.

To date, Eulerian methods (e.g. classical finite-difference, -volume, or -element methods) are most commonly used.
These employ a spatial grid, on which chemical species move in accordance with discretized forms of mass balance laws, such that they approximate transport governed by advection and dispersion.
To simulate chemical reactions, each grid point is treated as a well-mixed volume, and reactions are calculated based on the average concentrations of species residing within that volume.
Advantages of Eulerian methods include their intuitive nature, (relative) ease of implementation/parallelization, and the large body of mathematical research and justification supporting them.
This has led to their widespread use in various industrial and research applications \cite{TOUGHREACT,CRUNCHFLOW,PHREEQC,PHT3D}.

However, such methods also suffer from a variety of important drawbacks, such as the introduction of spurious numerical diffusion in the simulation of advection \cite{Sweby1984,Leonard1991,Leveque2002} and an inability to naturally capture fluctuations in concentration or mixing below scales resolved by the numerical grid.
As a result, if the system of interest requires capturing small-scale fluctuations, which can be important in the context of reactions \cite{battiato1,battiato2}, a very finely-discretized grid is required and may lead to high computational overhead, particularly as numerical stability conditions, dictated by advective velocity or dispersive/diffusive strength, impose restrictive time steps (e.g. $\dt / \dx^2 < c$ for some number $c$).
Alternatively, upscaling and the inclusion of additional closure terms that account for subscale effects can be included in the governing equations, but these too present significant challenges and restrictions depending on the complexity of reactions and competition between transport and reaction time scales \cite{battiato1,battiato2,Schwede_sample_pdf,Porta_2016}.
While empirical adjustments to reaction rates can often lead to better agreement between measurements and models, the physical basis for these adjustments (calibrations) can be questionable and typically cannot be scaled to other systems of interest.
Agreement is often only obtained through unphysical calibration and tweaking of model parameters, which works for hind-casting and observation fitting, but highlights the unphysical basis of many of these models and reveals problems with their use in a truly predictive sense.

Lagrangian methods, often referred to as particle-tracking (PT), do not employ a static spatial grid but rather discretize mass (or concentrations) into numerical ``particles,'' whose locations evolve in time, again following rules designed to capture advection and dispersion processes.
Generally speaking, one can define three sub-classes of these methods, grouped here according to how they simulate dispersion, typically dictated by the needs of the model.
One group simulates dispersion using random walks alone \cite{benson2008,Paster_WRR,Paster_JCP,Benson_AWR_2016,Bolster_mass,dong_awr,Ding_monod,Ding_WRR,Bolster_mass,schmidt2017,guillem2017kde,guillem_adaptive2018,guillem_grid_project19}.
A second simulates dispersion via mass-transfer between and among particles that do not random-walk and whose positions can only change by advection \cite{herrera_2009,herrera_2013,mass_trans_acc}.
The third group combines the random-walk and mass-transfer approaches \cite{Engdahl_WRR,herrera_2017,Benson_arbitrary,Schmidt_fluid_solid,guillem_SPH_equiv}, providing these algorithms the flexibility to model the distinct processes of mixing and spreading separately.
For methods that simulate dispersion via mass-transfer, there are two subdivisions whose equivalence, under specific modeling assumptions, was recently shown \cite{guillem_SPH_equiv}: smoothed-particle hydrodynamics (SPH) methods \cite{herrera_2009,herrera_2013,Gingold_originalSPH,Monaghan_SPHappl} and mass-transfer particle-tracking (MTPT) methods \cite{mass_trans_acc,Engdahl_WRR,Benson_arbitrary,Schmidt_fluid_solid,engdahl_ddc}.
In this work, we focus on the group of hybrid PT methods that use both random-walks and mass-transfer to simulate dispersion.

Some key advantages of particle-tracking methods include a lack of numerical diffusion when simulating advection and a natural ability to model arbitrarily steep concentration gradients, and thus the incomplete mixing inherent to many natural systems.
Because particle positions evolve continuously with time, regions of heterogeneity can evolve, move, and change size with time with arbitrarily fine resolution, so there is no ``homogeneity cutoff,'' as exists at the grid-scale of an Eulerian method.
Also, recent work has allowed for the parallelization of these algorithms, significantly reducing computational times \cite{engdahl_ddc,rizzo_GPU_PT}.
However, important drawbacks also arise from these methods.
First, the body of literature supporting PT methods is still relatively young, and there are open questions relating to the optimal particle number or time step length utilized in simulations \citep{schmidt2017}.
Additionally, while it has been conjectured, empirically demonstrated, and semi-analytically confirmed that these particle methods are simulating a perturbed reactive transport system \cite{Paster_JCP,schmidt2017}, it is not clearly understood how different methods of perturbing the simulation correspond to observable heterogeneities in the real world.
To date, these approaches have only been applied to relatively simple reactive systems, consisting of a somewhat small number of reactive components and reactions.

In order to address these challenges, we apply the mobile-immobile reactive particle-tracking (miRPT) model of \cite{Schmidt_fluid_solid} that allows for interaction between aqueous (mobile) solutes and immobile mineral phases to simulate a chemically-complex benchmark system.
In the course of this application, we examine the effects of a key modeling choice that is inherent to the miRPT algorithm--the choice of how to represent solid species via immobile particles.
Specifically, what happens if we perturb this representation, and what is the meaning of different types of perturbations, as it relates to reactant inhomogeneity or imperfect mixing?
From a physical perspective, it seems intuitive that the spatial configuration of solid species should influence the results of a reactive transport simulation.
In order to capture a highly heterogeneous state, Eulerian methods would require an increased level of spatial discretization to resolve solid species distributions, driving up computation times and imposing stricter stability conditions.
In fact, due to the computational burden imposed by such conditions, the family of particle-tracking models considered in this work tend to display run times significantly lower than first-order Eulerian methods and less than half that of a more accurate 3$^{\text{rd}}$-order method
\cite{Benson_AWR_2016}.

Particle-tracking methods also offer increased flexibility in their ability to capture heterogeneity.
For a sufficient number of immobile particles, a user may distribute them in space in a fully-controllable manner.
This may be done to fit a desired continuous distribution, incurring no extra computational cost, as no increase in discretization is required.
Also, due to the local nature of particle methods, in that particles that are not near one another do not interact via mass-transfer, they are able to capture poor mixing in a way that Eulerian models cannot.
Additionally, without increasing model complexity, the user may run an ensemble of such simulations in order to inform descriptive statistics about the system, rather than just obtaining point estimates.

The test problem we use to investigate the above-mentioned issues is presented in \cite{Arora2015}, wherein the authors consider the problem of heavy metal cycling in lake sediments (henceforth referred to as the HMLS system).
The authors use the system as a benchmark to compare several popular and state-of-the-art Eulerian reactive transport models (TOUGHREACT \cite{TOUGHREACT}, CrunchFlow \cite{CRUNCHFLOW}, PHREEQC \cite{PHREEQC,phreeqcrm}, and PHT3D \cite{PHT3D}).
A selection of the results of this benchmark study are shown in Figure \ref{fig:arora_fig3_and_4}, in which their results are compared to experimental data from \cite{winowiecki_thesis}.
Note that while all of the Eulerian models yield nearly identical results, none of them capture the variability in the data, nor do they capture visible fluctuations of certain species.
This is most evident in Figure \ref{fig:arora_fig3_and_4}(f), depicting Pb$^{\text{+2}}$, where we see an order-of-magnitude difference between the two data plots and a non-smooth distribution of the data in space.
Neither of these is captured by the single-solution smooth curves provided by the Eulerian models, but this variability is exactly the type of behavior that can be captured by our stochastic particle-tracking model, especially when using an ensemble of realizations employing perturbed conditions.

We organize our investigation of this benchmark reactive transport system as follows.
In Section \ref{sec:governing_model}, we discuss the mathematical model and describe relevant physical processes.
In Section \ref{sec:numerical_implementation}, we describe how a finite-difference method, later used as a base-case for comparison with our PT models, is implemented and provide details germane to our implementation of the miRPT algorithm.
In Section \ref{sec:results_Hmetal}, we outline the results of simulating the HMLS system with the miRPT algorithm, where we develop unperturbed and perturbed models.
Finally, conclusions are presented in Section \ref{sec:conclusions_hMetal}.


\section{Governing model} 
\label{sec:governing_model}

\subsection{Conceptual description} 
\label{sub:conceptual_description}

\begin{figure}[t]
    \centering
    \includegraphics[width=1.0\textwidth]{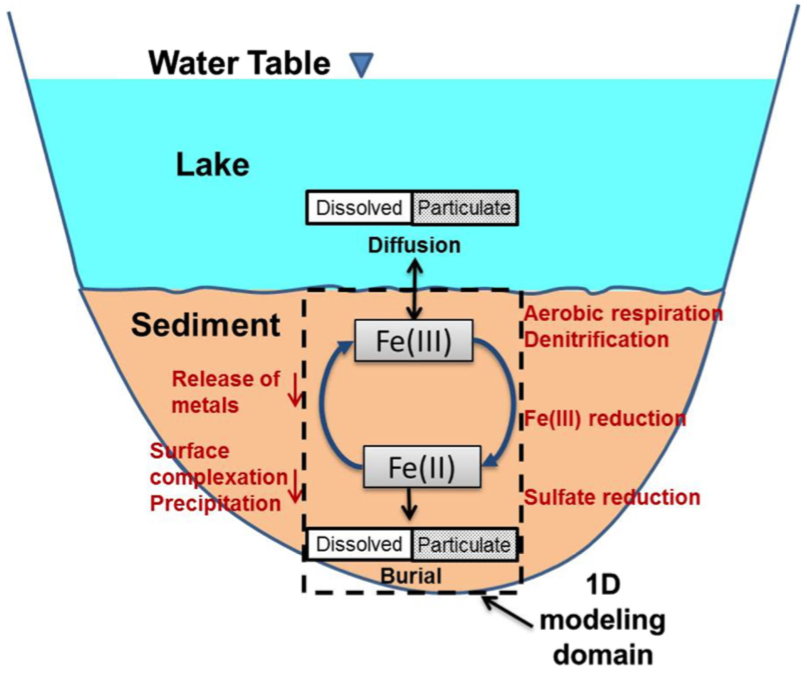}
    \caption{Figure 1 from \cite{Arora2015}, used with permission. Schematic diagram depicting the modeling domain and biogeochemical processes governing the heavy metal cycling in lake sediments (HMLS) system.}
    \label{fig:arora_schematic}
\end{figure}

For brevity, we refer the reader to the work of \emph{Arora et al.} \cite{Arora2015} for a highly detailed description of the system and chemical reactions involved (see Figure \ref{fig:arora_schematic} for their schematic diagram).
Here we provide only a brief synopsis of the principal physico-chemical processes.
In general, the system represents a transition from oxic conditions at an upper lake water boundary to anoxic, or reducing conditions, deeper in the sediment column.
Oxygenated water with an electron donor (food source) of acetate is present in the lake water and enters the column from the top.
Microbial aerobic respiration converts the acetate to carbonate ions (represented here by total alkalinity) and reduces aqueous-phase electron acceptors (in order of preference or ease of conversion) O$_2$, nitrate $\left(\text{NO}_3^{-}\right)$, Fe$^{+3}$, and sulfate $\left(\text{SO}_4^{-2}\right)$.
Reductive dissolution of solid-phase (in the lake sediments) ferrihydrite (Fe(OH)$_3$) in this sequence also releases metals sorbed to the mineral surface, including lead and zinc.
These dissolved metals and associated aqueous complexes may diffuse back into the upper-boundary lake water, or react with biogenically produced sulfide from sulfate reduction in the lake sediments and precipitate as solid sulfide minerals.


\subsection{Mathematical formulation} 
\label{sub:mathematical_formulation}

The mathematical model we consider is governed by the 1D diffusion-reaction equation (DRE)
\begin{equation}\label{hMetal_DRE}
    \begin{aligned}
        \pder{t}\left(\phi C_i\right) &= D \Delta C_i + r_i(C_1, \dots, C_n, k_1, \dots, k_m),\\
        i &= 1, \dots, n, \quad x \in \Omega \subseteq \R, \quad t > 0,
    \end{aligned}
\end{equation}
where $C_i(t, x)$ [mol m$^{-3}$] is the concentration of chemical species $i$, $\phi$ [dimensionless] is porosity, $D$ [m$^2$ s$^{-1}$] is the diffusion coefficient (defined to be constant and equal for all aqueous species and zero for solid species), and $r_i$ [mol m$^{-3}$ s$^{-1}$] is a source/sink term representing chemical reaction that, for each species $i$, is a function of some number of the $n$ species and $m$ reaction coefficients, $k_j$.
The system that we focus on is referred to as the ``biotic case'' or ``base case'' in the work of \emph{Arora et al.} \cite{Arora2015}.
Their schematic diagram is reproduced in Figure \ref{fig:arora_schematic}, but we also summarize here.
In this system, the modeling domain, $\Omega$, is a vertically-oriented sediment column of length $L = 0.4$ m in the downward direction, i.e., $\Omega = [0.0, -0.4]$.
The initial condition (IC) is constant throughout the domain and consists of a specific distribution of the species of chemical reactants \cite[see][Table 6]{Arora2015}.
The boundary conditions (BCs) are Dirichlet (constant and equivalent to the initial condition) at the upper boundary ($x = 0$), or sediment/lake-water interface, and Neumann (zero flux) at the lower boundary ($x = -L$), which represents the lake bed, such that
\begin{equation}\label{icbc}
    \begin{aligned}
        C_i(t = 0, x) &= C_i^0,\\
        C_i(t, x = 0) &= C_i^0,\qquad\qquad i = 1, \dots, n,\\
        \pder{x}C_i(t, x = -L) &= 0.
    \end{aligned}
\end{equation}
A table showing the initial concentrations, $C_i^0$, for key chemical species of interest is given in Table \ref{tab:hMetal_ICBC}; note that the initial pore water chemistry has not yet been allowed to equilibrate with the mineral phases in the sediments.

\begin{table}[t]
    \centering

    \begin{tabular}{|c|l|}
    \hline
    \multicolumn{2}{|c|}{\textbf{Concentration}} \\
    \hline
    pH & 7.2 \\
    \hline
    Alkalinity  & $0.72 \times 10^{-2\phantom{2}}$ mol/kgw \\
    \hline
    NO$_3^-$    & $0.80 \times 10^{-5\phantom{2}}$ mol/kgw \\
    \hline
    SO$_4^{-2}$ & $0.58 \times 10^{-4\phantom{2}}$ mol/kgw \\
    \hline
    Fe$^{+2}$   & $0.48 \times 10^{-12}$ mol/kgw \\
    \hline
    Pb          & $0.55 \times 10^{-7\phantom{2}}$ mol/kgw \\
    \hline
    \end{tabular}
    \caption{Initial/boundary concentrations of selected chemical species, corresponding to those shown in the figures of Section \ref{sec:results_Hmetal}.}
     \label{tab:hMetal_ICBC}
\end{table}

As to the values of relevant model parameters, we use those found in \cite[][Table 5]{Arora2015}, namely $\phi = 0.47$, $D = 4.27 \times 10^{-10}$ m$^2$ s$^{-1}$, and a total simulation time of $T = 5$ years.
We clarify here that based on the description in \cite{Arora2015} of the problem and correspondence of our numerical results (Section \ref{sec:results_Hmetal}) with theirs, holding porosity constant in time appears to be the appropriate modeling choice in this case.
As such, we absorb $\phi$ into $D$ (and the reactive term) for an effective diffusion coefficient of $D^* \defeq D / \phi \approx 9.09 \times 10^{-10}$ m$^2$ s$^{-1}$.
Chemical reaction parameters are not discussed here, as they are defined within a PHREEQC input file and database that were provided by the authors of \cite{Arora2015} so as to match those used in their simulations.
These files, along with the rest of the code used to generate the results in Section \ref{sec:results_Hmetal}, are provided in the following repository
\begin{center}
\url{https://github.com/mschmidt271/metalsCode_thesis}. 
\end{center}



\section{Numerical implementation} 
\label{sec:numerical_implementation}

In order to re-create the results of \cite{Arora2015} and consider the effect of spatial perturbations in the PT model, we consider two numerical approaches: a finite-difference (FD) model that we use as a base-case for comparison and one employing the miPRT algorithm \citep{Schmidt_fluid_solid}.
Both models use the phreeqcRM reaction module \cite{phreeqcrm} for chemistry calculations, which is driven by the PHREEQC input file and database provided by the authors of \cite{Arora2015}.

\subsection{Finite-difference model} 
\label{sub:finite_difference_model}

The FD model we use as a base case is on a regularly-spaced grid, explicit in time, and second-order, centered in space; we use an operator-splitting approach between diffusion and reaction calculations.
We choose a spatial grid with spatial step size $\dx = 1$ cm (implying a number of cells, $N_C \defeq L / \dx = 40$), in contrast to \cite{Arora2015}, which uses 46 cells, spaced by 0.5 cm for the top 8 and by 1 cm for the rest of the domain.
We consider three different time step lengths of $\dt \in \{259, 2592, 25920\}$ s, so as to calibrate our model by exploring the range of time step lengths explored by \cite{Arora2015}, while obeying the von Neumann stability condition
\begin{equation*}
    \frac{D^* \dt}{\dx^2} \leq \frac{1}{2}.
\end{equation*}


\subsection{Particle-tracking model} 
\label{sub:particle_tracking_model}

The PT model we use employs the miRPT algorithm of \cite{Schmidt_fluid_solid}.
That algorithm is based on \cite{Bolster_mass} which reformulated reactive PT algorithms in terms of mass reduction, rather than a particle-killing approach.
The work of \cite{Benson_arbitrary} extended this algorithm such that particles could carry an arbitrary number of chemical species that are transferred among particles via diffusive mass transfer, and \cite{Engdahl_WRR} added phreeqcRM \cite{phreeqcrm} to handle complex geochemical reactions between the species.
The algorithm of \cite{Schmidt_fluid_solid} further added the capability for fluid-solid (mobile-immobile) interactions.

For this implementation, we partition the total (effective) diffusion of the system such that
\begin{equation}
    D^* \defeq D_{\text{RW}} + D_{\text{MI}} + D_{\text{IM}},
\end{equation}
where $D_{\text{RW}}$, $D_{\text{MI}}$, and $D_{\text{IM}}$ are the portions of the total diffusion simulated by random-walks, mobile-to-immobile mass-transfers, and immobile-to-mobile mass-transfers, respectively; we impose values of $D^* \times \{0.5, 0.25, 0.25\}$.
This means that half of the total diffusion in the system is simulated by random-walking mobile particles, and the remaining half is simulated by the two ``directions'' of mass-transfer.
For the purposes of this project, these values were chosen ad-hoc, though it has been suggested that properly calibrating this partitioning allows for separate simulation of the distinct processes of mixing and spreading \cite{guillem_SPH_equiv,mass_trans_acc}.
For a discussion of the effect of these modeling choices, see Appendix A in \cite{Schmidt_fluid_solid}.
The time step lengths that we employ are chosen so as to correspond to the results given by the finite-difference simulations, and so we consider $\dt \in \{259, 2592, 25920\}$ s.

The number of mobile particles ($N_M$) is held to be 4000 for all simulations, though we consider different numbers of immobile particles ($N_I$), depending on the effects we wish to examine.
For the base (unperturbed) case we consider $N_I \in \{40, 100, 400\}$, and for the perturbed cases we consider $N_I \in \{100, 400\}$.
The reason for the disparity between $N_I$ and $N_M$ is that we wish to highly resolve spatial gradients with a large number or mobile particles because transport calculations are computationally cheap; however, because the highly-expensive chemistry calculations are conducted on the immobile particles, we would like to minimize $N_I$ to the smallest appropriate level.
To put numbers to this concept, when employing $(N_I, N_M) = (40, 4000)$ and running on a laptop machine, the chemistry calculations are $\cO(100)$ times more expensive than all of the transport calculations within a time step, despite the chemistry calculations being conducted in parallel on 4 cores and the transport being conducted in serial.
Finally, the zero-flux Neumann condition at $x = -L$ is enforced as a reflecting boundary \cite{Szymczak_particle_BCs}.

\subsubsection{Reformulated optimality condition} 
\label{ssub:reformulated_opt_cond}

Here, we provide a brief discussion of the ``optimality condition'' for the miRPT algorithm \cite{Schmidt_fluid_solid}. In particular, the simulation constraint
\begin{equation}\label{bad_eta}
    \eta \defeq \frac{(L/\min(N_I, N_M))^2}{D^* \Delta t} \leq 1,
\end{equation}
attempts to ensure that the maximum average inter-particle spacing (whether of mobile or immobile particles) is of the same order as the magnitude of the system's diffusion within a time step of length $\dt$.
As such, there should always be a ``nearby'' mobile or immobile particle to receive mass-transfers from a given particle of the opposite type.
Violation of this condition does not necessarily lead to unstable solutions that ``blow up,'' as occurs when one violates a von Neumann stability condition in a finite-difference simulation.
Instead, the accuracy of the solution degrades in a relatively steady manner as we move further from the prescribed optimal values of $\eta$.

We acknowledge that many of the previously mentioned choices of discretization parameters, $N_I,\ N_M,\ \dt$ violate this condition.
In \cite{Schmidt_fluid_solid}, $N_I$ and $N_M$ were always close to the same order of magnitude, while here this is not feasible due to the highly expensive chemistry calculations (i.e., we always choose to have fewer immobile particles, in order to minimize the cost of these chemistry calculations).
As a result, a more nuanced formulation of this condition is presented for our purposes.

We would like to enforce a condition such that, for a given transfer (denoted ``MI'' for mobile-to-immobile and vice versa ``IM''), the expected distance between two particles of opposite species is less than the standard diffusion distance $\ell \defeq \sqrt{2 D \dt}$.
This distance is maximized when a particle is located directly between two particles of the opposite species, so in the case of an MI transfer, the maximum distance of a mobile particle to the nearest immobile particle is $L / (2 N_I)$.
Thus, we would reformulate the individual optimality conditions for a given MI or IM transfer as
$$ \frac{\frac{L}{2 N_I}}{\sqrt{2 D^* \dt}} \leq 1 \qquad \mathrm{and} \qquad \frac{\frac{L}{2 N_M}}{\sqrt{2 D^* \dt}} \leq 1$$
or upon simplifying expressions
\begin{equation}
    \begin{aligned}
        \eta_{\text{\ MI}} &= \frac{\left(\frac{L}{N_I}\right)^2}{D^* \dt} \leq 8,\\
        \eta_{\text{\ IM}} &= \frac{\left(\frac{L}{N_M}\right)^2}{D^* \dt} \leq 8.
    \end{aligned}
\end{equation}
Hence, we choose all values of $N_I,\ N_M,$ and $\dt$ in the following sections so as to satisfy these conditions.

\section{Results} 
\label{sec:results_Hmetal}

Next, we examine the effect of various modeling choices on the results of simulating the HMLS system via the miRPT algorithm.
All plots in this section depict final time results (5 years) within the domain for relevant chemical species (and pH), corresponding to those from Figures 3 and 4 of \emph{Arora et al.} \cite{Arora2015}, which are reproduced in Figure \ref{fig:arora_fig3_and_4} for reference.
The plots that correspond to \cite[][Figure 3]{Arora2015} depict final concentrations of key chemical markers, including important aqueous ions, pH, and alkalinity, and they aim to recreate experimental data from \cite{winowiecki_thesis}.
We obtained a copy of the Winowiecki thesis \cite{winowiecki_thesis}, which includes the data that is used by \cite{Arora2015}.
The data in the thesis include two trials at two different sampling times, summer and fall of 2001, and we portray all of this data in order to highlight its variability (note that the two data scatters in \cite{Arora2015} depict summer, trial 2 and fall, trial 1).
In Figures \ref{fig:hMetal_FD_rangeDT_3x3}, \ref{fig:hMetal_Ni100_dt25920_rand_concs_ens_3x3}, and \ref{fig:hMetal_Ni100_dt25920_ens_3x3}, we plot the data for pH, sulfate, Fe$^{+2}$, and Pb (plots (a) and (d)-(f), respectively); note however, that there are not two trials for each sampling for pH, so only two data scatters are shown in plot (a).
The plots that correspond to \cite[][Figure 4]{Arora2015} depict the percentage difference between initial and final amounts of three secondary iron species for which there is not data but whose behavior would be affected by the presence of the previously-mentioned aqueous ions.
All of the code used to generate the results in this Section, including the data from \cite{winowiecki_thesis}, may found at
\begin{center}
\url{https://github.com/mschmidt271/metalsCode_thesis}.
\end{center}

\subsection{Verification} 
\label{sub:verification}

\begin{figure}[h]%
    \centering
    \includegraphics[width=1\textwidth]{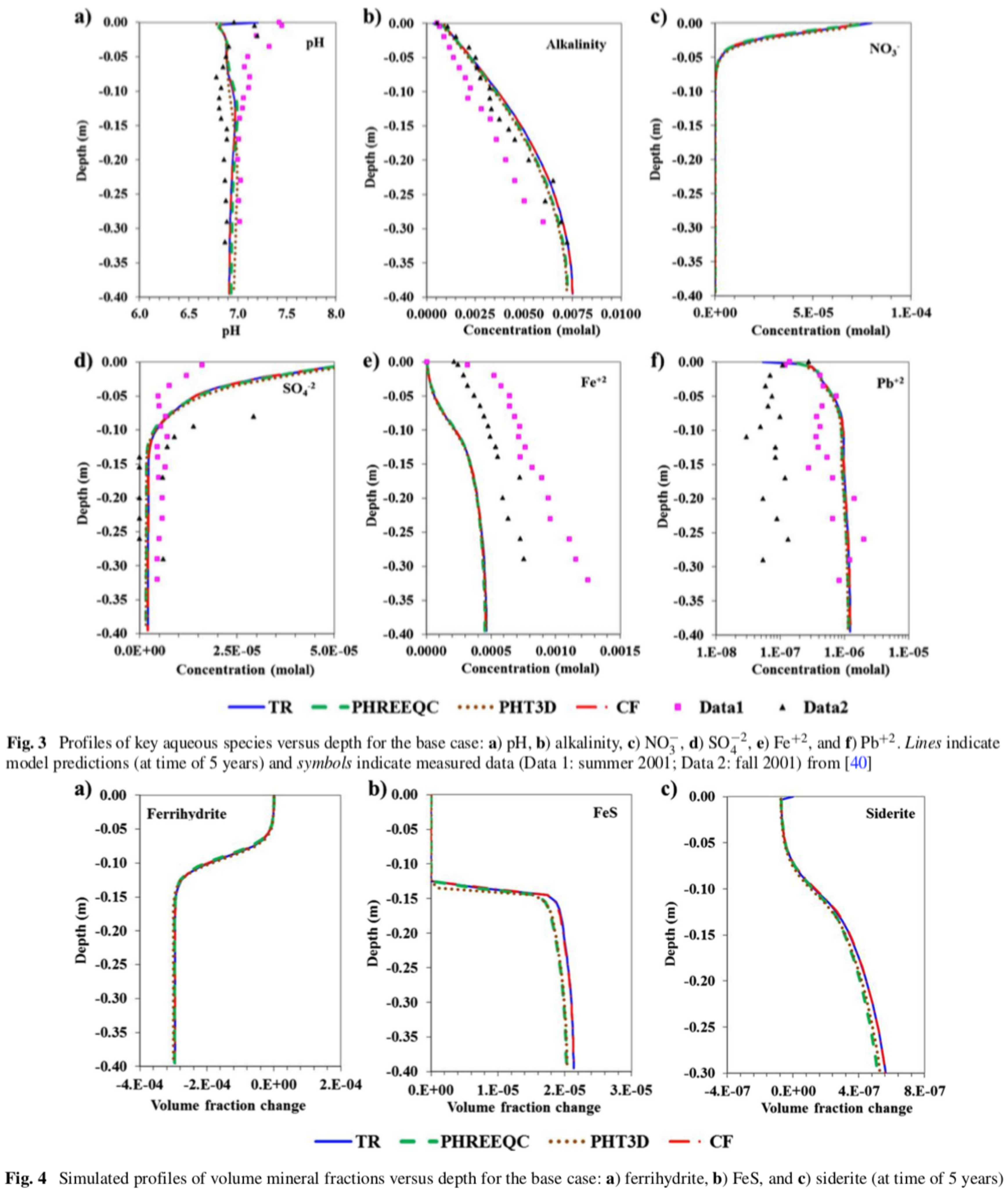}
    \caption{(Top) Figure 3 from \emph{Arora et al.} \cite{Arora2015}, depicting key chemical markers.
    Note that, while the bottom right plot is labeled Pb$^{+2}$, this should be Pb (See \cite{sengor}, Figure 7(c) to confirm this).
    As well, while (b) is labeled as ``alkalinity,'' \cite{sengor} refer to this quantity as ``bicarbonate alkalinity,'' and for this reason, we plot actual alkalinity in all of our subsequent plots, leading to the apparent discrepancy between our plots and (b), here.
    (Bottom) Figure 4 from \emph{Arora et al.} \cite{Arora2015}.
    The percentage difference between the initial and final amount of the secondary iron species, after 5 years of simulation time, is depicted. Used with permission.}
    \label{fig:arora_fig3_and_4}
\end{figure}

\begin{figure}[h]%
    \centering
    \includegraphics[width=1\textwidth]{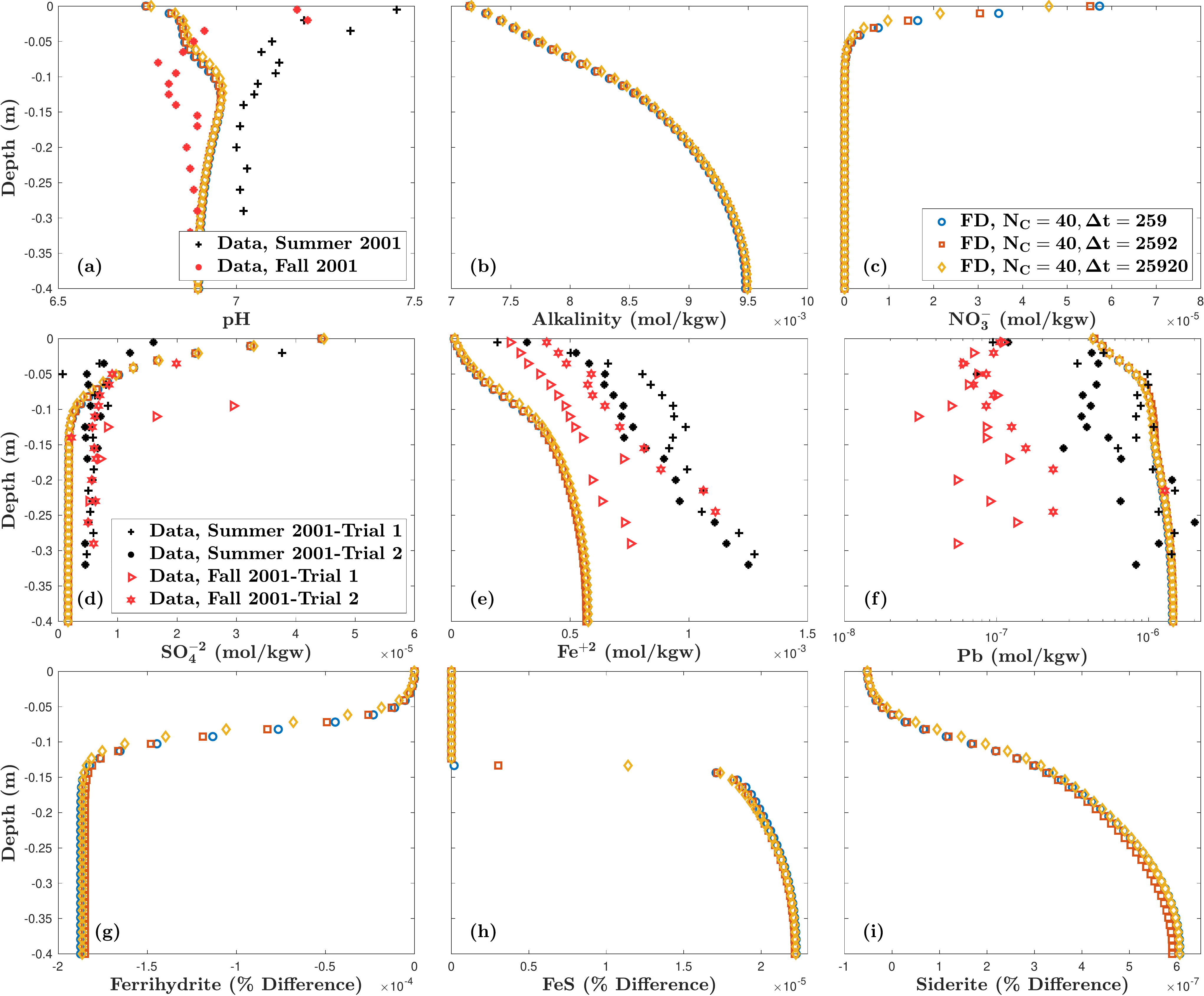}
    \caption{FD base-case simulations with 40 cells for varying values of $\dt$.
             Final time concentrations ($t = 5$ years) of key chemical markers and pH are shown in plots (a)-(f) and percentage difference between initial and final amounts of secondary iron species is shown in plots (g)-(i).
             Data from \cite{winowiecki_thesis} is plotted against simulated results for (a), (d)-(f).}
    \label{fig:hMetal_FD_rangeDT_3x3}
\end{figure}

We first verify that the results of \emph{Arora et al.} \cite{Arora2015} can be recreated by implementing the FD simulation described in Section \ref{sub:finite_difference_model}.
The results of running FD simulations for values of $\dt$ ranging three orders of magnitude (corresponding to the time steps used in the Eulerian models of \cite{Arora2015}) are displayed in Figure \ref{fig:hMetal_FD_rangeDT_3x3} and show that, for this model, the length of the time step does not appear to result in a significant difference in the results.
All plots shown in Figure \ref{fig:hMetal_FD_rangeDT_3x3} demonstrate nearly identical behavior and correspond closely to those shown in Figure 3 of \cite{Arora2015} (reproduced here in Figure \ref{fig:arora_fig3_and_4}).
One noticeable difference, however, is the appearance of slight changes in the upper $(x = 0)$ boundary concentrations of nitrate (NO$_3^-$, Figure \ref{fig:hMetal_FD_rangeDT_3x3}(c)) related to the chemistry calculations performed by phreeqcRM.
Such deviations occur, as the authors of \cite{Arora2015} note in their provided PHREEQC input file, because equilibrium is never reached for nitrate under the conditions they consider.
This is evidenced by the discrepancy between initial/boundary concentrations and concentrations within the domain at the $x = 0$ boundary, the magnitude of which appears to be highly dependent on time step length.
Thus, a longer time step results in more consumption of the nitrate entering at the upper boundary, explaining the lower concentrations as the time step is increased.
The authors also note the same for sulfate (SO$_4^{2-}$, Figure \ref{fig:hMetal_FD_rangeDT_3x3}(d)), though the effects are less dramatic in these FD cases.
The results shown in Figure \ref{fig:hMetal_FD_rangeDT_3x3} indicate that little error is induced in the chemistry calculations by choosing a time step in the given range, and, provided that it does not introduce error in the transport calculations, we may consider time step lengths in this range for PT simulations.


\subsection{Particle-tracking results for base-case model} 
\label{sub:particle_tracking_results_base}

\begin{figure}[h]%
    \centering
    \includegraphics[width=1\textwidth]{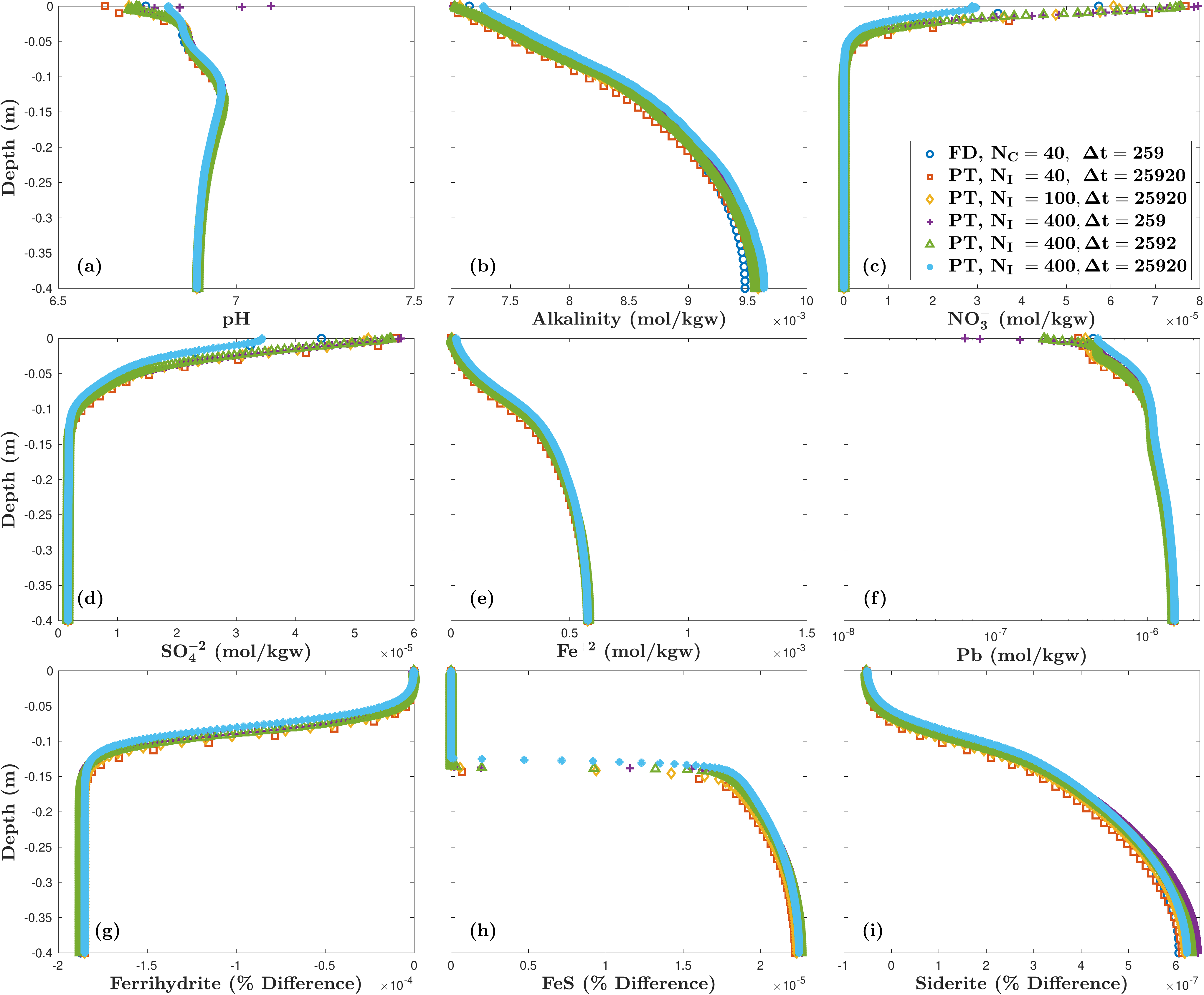}
    \caption{Comparison of base-case FD simulation to 40 and 400 equally-spaced immobile particle simulations, for various values of $\dt$.
    Final time concentrations ($t = 5$ years) of key chemical markers and pH are shown in plots (a)-(f) and percentage difference between initial and final amounts of secondary iron species is shown in plots (g)-(i).}
    \label{fig:hMetal_FD_evenPT_3x3}
\end{figure}

Now that the FD simulations have been faithfully reproduced, we will use the miRPT algorithm to model the HMLS system.
We first verify that the prior results are produced in the case of equally-spaced immobile particles.
Figure \ref{fig:hMetal_FD_evenPT_3x3} displays the results of PT simulations for immobile particles that are equally-spaced across the domain and possess the same initial concentrations as in the FD case.
We compare these to the finest time-discretization of the FD results, though they were similar for all tested values of $\dt$.
We find close agreement between most of the PT simulations and the FD results, with the cases of $(N_I, \dt) = (\{40, 100\}, 25920)$ and $(N_I, \dt) = (400, \{259, 2592\})$ all displaying nearly identical behavior.
These PT simulations diverge from the FD results in the case of nitrate and sulfate concentrations (NO$_3^-$ and SO$_4^{2-}$, Figure \ref{fig:hMetal_FD_evenPT_3x3}(c) and (d), respectively). However, this difference is attributable to the effect of the time step on the chemistry calculations, as described in Section \ref{sub:verification}.

The other differences occur in the alkalinity (Figure \ref{fig:hMetal_FD_evenPT_3x3}(b)), but are relatively small.
We note that the $N_I = 400$ simulations seem to resolve a sharper gradient in Pb concentration at the upper boundary $(x = 0)$ than the FD or $N_I = \{40, 100\}$ cases, with a smaller time step corresponding to a sharper apparent gradient.
The one simulation that shows the most significant difference from the other PT simulations is the $(N_I, \dt) = (400, 25920)$ case, though these differences are still minor, both in comparison to the FD simulation and the remaining PT simulations.
Overall, the PT and FD models yield highly comparable results.


\subsection{Perturbation analysis for particle-tracking model} 
\label{sub:perturbation_analysis}

Next, we investigate the miRPT model described in Section \ref{sub:particle_tracking_model} under two different perturbation approaches.
We first study the effects of perturbing the initial aqueous-phase concentrations stored on the evenly-spaced immobile particles (Section \ref{ssub:IC_perturbations}), and later study perturbations of the spatial locations of the immobile particles (Section \ref{ssub:randomly_spaced_immobile_particles}).
The former approach introduces a level of variability into the initial distribution of reactants without affecting the overall mixedness of the system (i.e., the level of mixing by interaction with mobile species remains the same as in the unperturbed case).
Thus, these results should occupy a middle ground between the unperturbed case and the spatially-perturbed case in the subsequent section.
Contrastingly, the latter approach is meant to physically represent an irregular spatial distribution of solid species, which can also be thought of as introducing a poorly-mixed condition to the system.
These spatial gaps between reactants will take time for aqueous (mobile) species to traverse, delaying their contact with other reactive species.
The associated increase in travel time will lead to a slowdown of reaction speed that should be apparent in the shape and position of reaction fronts.
We note that initial concentration perturbation, considered in Section \ref{ssub:IC_perturbations}, is a method that can be achieved similarly by using Eulerian methods. However, the position perturbations, considered in Section \ref{ssub:randomly_spaced_immobile_particles}, are not so easily achieved with grid-based methods and represent a task to which particle-tracking methods are uniquely suited.
In both cases, we conduct an ensemble of 100 realizations in order to accurately capture the statistics of the results.

\subsubsection{Perturbations of immobile particle initial concentrations} 
\label{ssub:IC_perturbations}

\begin{figure}[h]%
    \centering
    \includegraphics[width=1\textwidth]{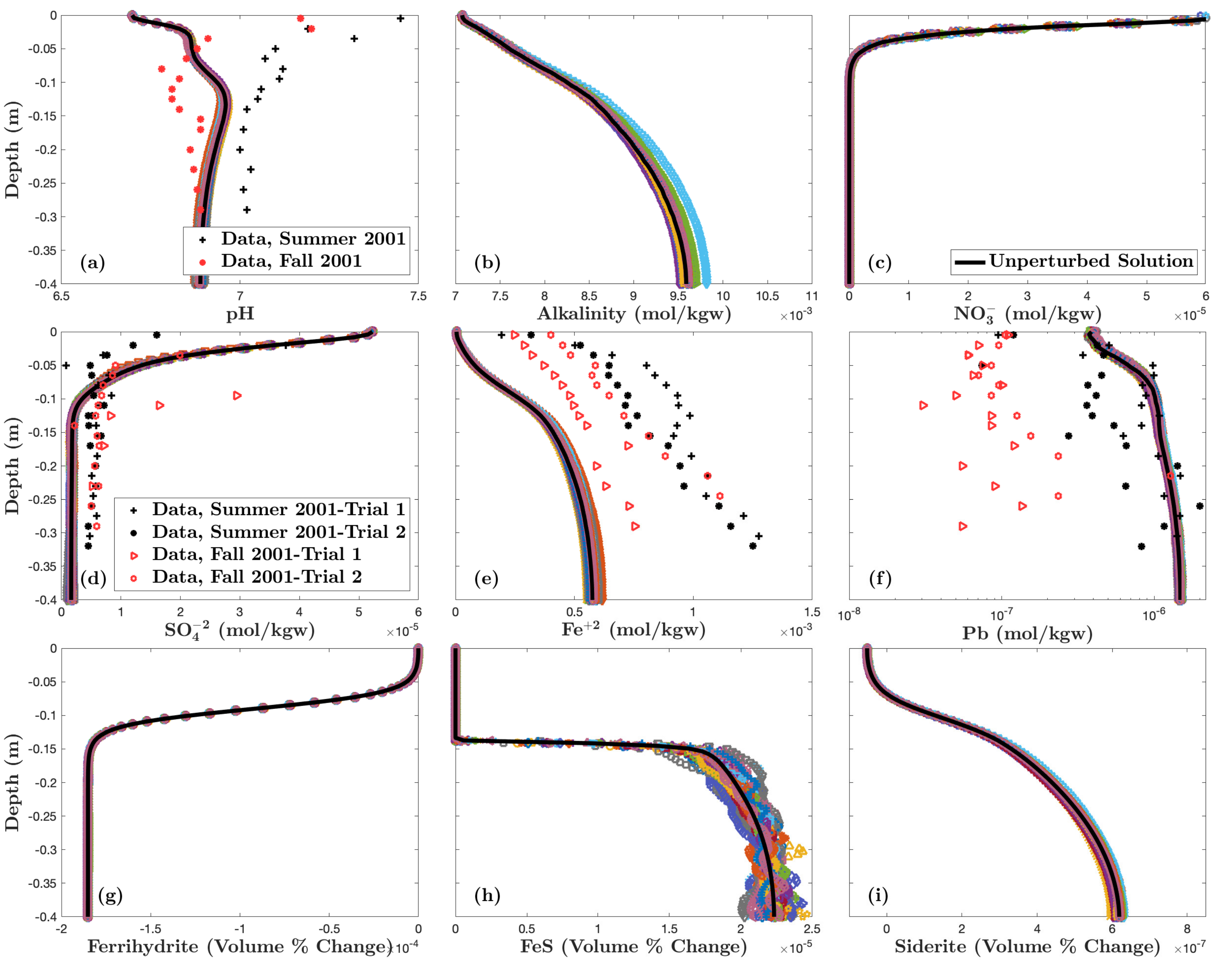}
    \caption{Ensemble results (100 realizations) for 100 equally-spaced immobile particles with randomly-perturbed initial concentrations and $\dt = 25920$.
    Results of individual simulations are depicted.
    Final time concentrations ($t = 5$ years) of key chemical markers and pH are shown in plots (a)-(f) and percentage difference between initial and final amounts of secondary iron species is shown in plots (g)-(i).
    Data from \cite{winowiecki_thesis} is plotted against simulated results for (a), (d)-(f).
    }
    \label{fig:hMetal_Ni100_dt25920_rand_concs_ens_3x3}
\end{figure}

\begin{figure}[h]%
    \centering
    \includegraphics[width=1\textwidth]{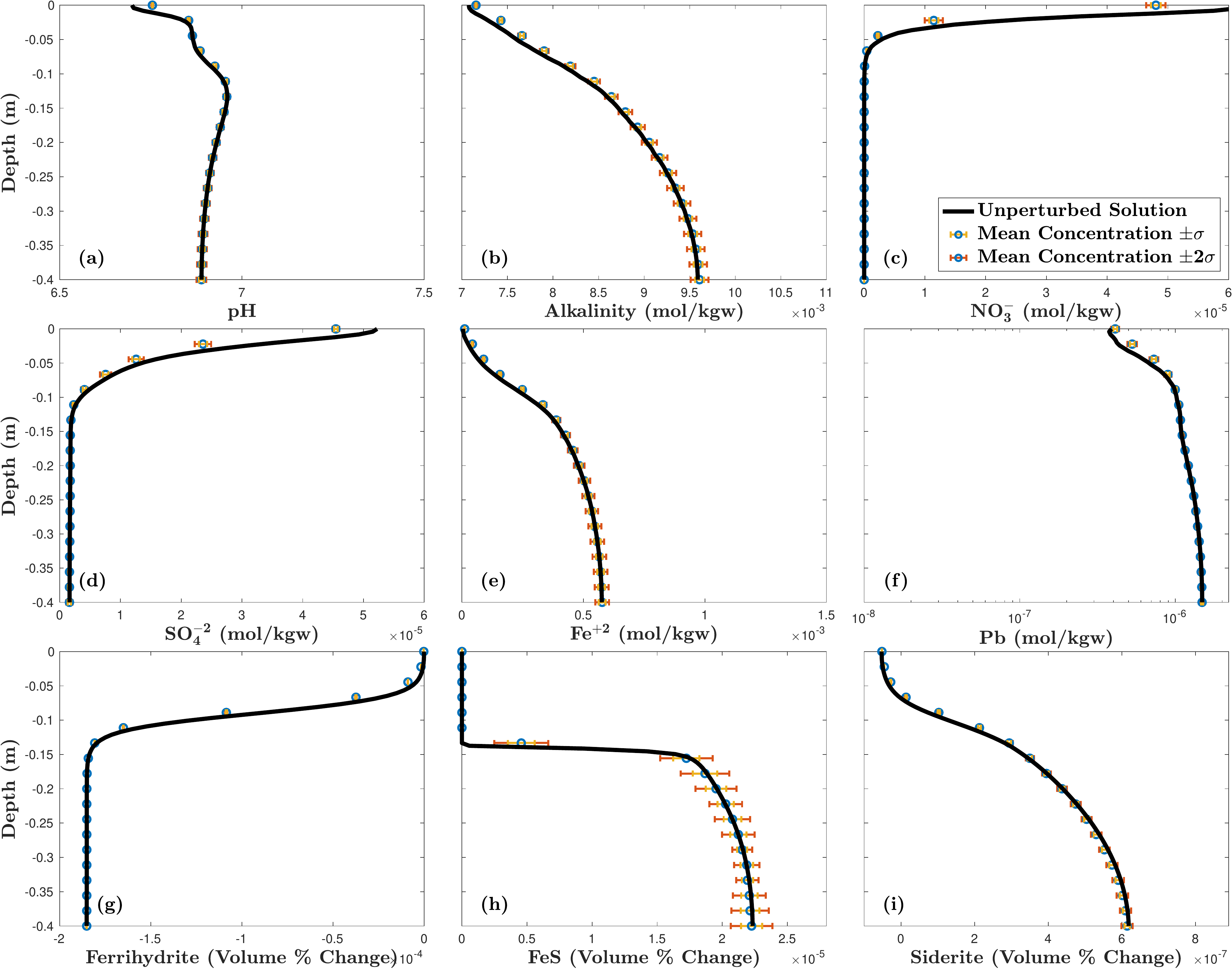}
    \caption{Ensemble results (100 realizations) for 100 equally-spaced immobile particles with randomly-perturbed initial concentrations and $\dt = 25920$.
    Final time concentrations ($t = 5$ years) of key chemical markers and pH are shown in plots (a)-(f) and percentage difference between initial and final amounts of secondary iron species is shown in plots (g)-(i).}
    \label{fig:hMetal_Ni100_dt25920_rand_concs_eBar_3x3}
\end{figure}

In this section, we employ equally-spaced immobile particles (as in Section \ref{sub:particle_tracking_results_base}), but uniformly perturb the initial concentrations that are stored on these particles.
This is achieved by choosing an amount, $\alpha \in (0, 1)$, and for each species $i$ and immobile particle $j$, perturbing the initial concentration $C_0^{ij}$ according to a draw from a Uniform, $\cU\left((1 - \alpha) C_0^{ij}, (1 + \alpha) C_0^{ij} \right)$, distribution.
In other words, we perturb the initial concentration by $\pm 100 \alpha \%$, and for the results of this section, we have selected $\alpha = 0.8$.
For this analysis, we run simulations with $(N_I, \dt) = (100, 25920)$ and $(N_I, \dt) = (400, \{2592, 25920\})$.
The results for these levels of discretization were similar enough that we only depict and discuss $(N_I, \dt) = (100, 25920)$.

Figure \ref{fig:hMetal_Ni100_dt25920_rand_concs_ens_3x3} shows the final time concentrations for each of the 100 realizations, and Figure \ref{fig:hMetal_Ni100_dt25920_rand_concs_eBar_3x3} shows the ensemble mean (blue markers) and $\pm 1$ and $\pm 2$ standard deviations of the concentration ensemble results (yellow and orange error bars, respectively).
In all plots, we compare to the unperturbed solution of Section \ref{sub:particle_tracking_results_base} (solid black line), and we also show the data from \cite{winowiecki_thesis} in selected plots.
So that we may compare to the results in the following section, in which the positions of immobile particles were randomly chosen and differ for each member of the ensemble, the ensemble mean and standard deviations were computed by binning concentrations into equally-spaced bins.
The bin size is chosen such that at least one particle is in each bin, and this resulted in 19 bins.

\subsubsection{Perturbations of immobile particle spatial positions} 
\label{ssub:randomly_spaced_immobile_particles}

\begin{figure}[h]%
    \centering
    \includegraphics[width=1\textwidth]{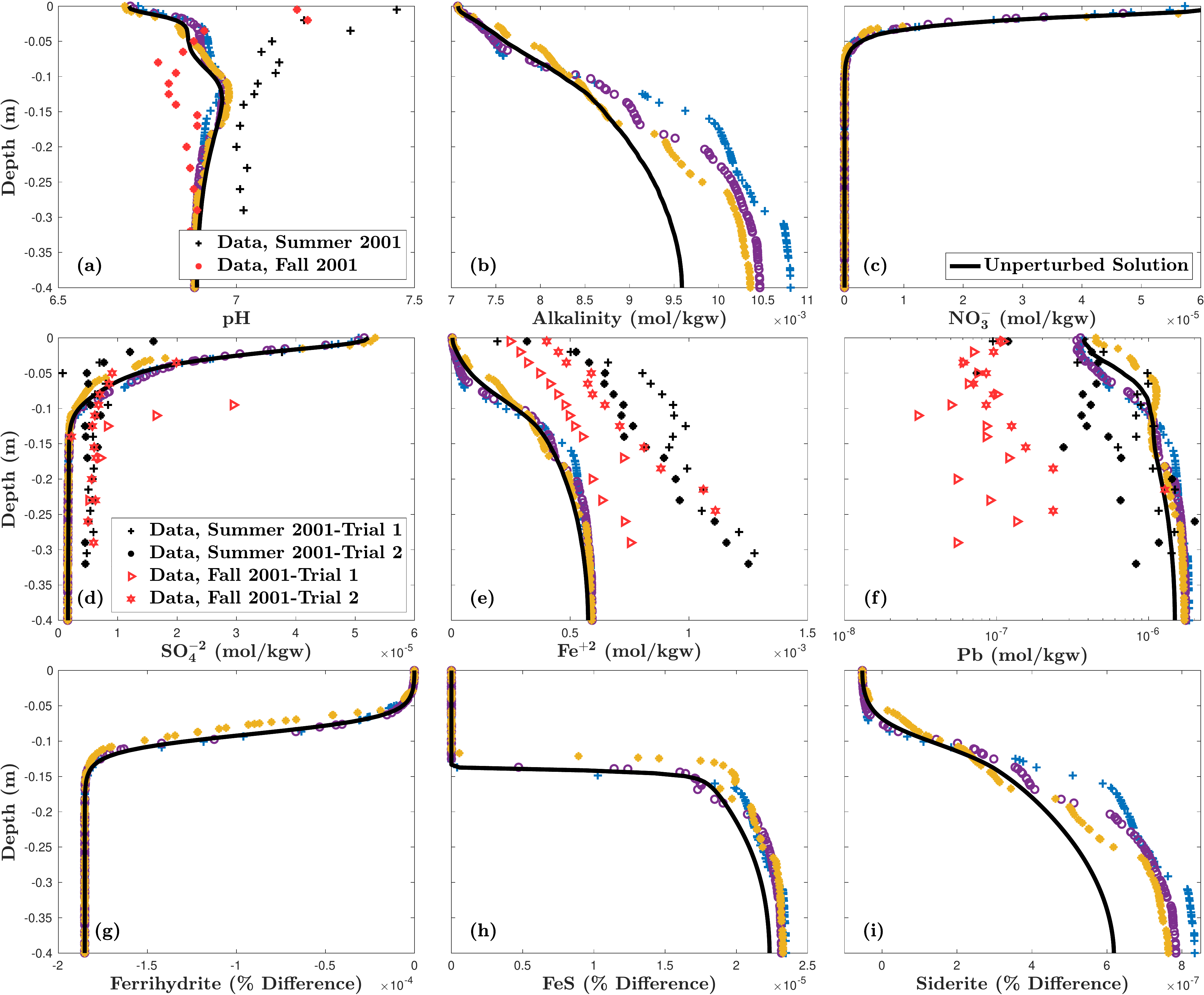}
    \caption{Selected ensemble results (3 out of 100 realizations) for 100 randomly-spaced immobile particles and $\dt = 25920$.
    Results of individual simulations are depicted.
    Final time concentrations ($t = 5$ years) of key chemical markers and pH are shown in plots (a)-(f) and percentage difference between initial and final amounts of secondary iron species is shown in plots (g)-(i).
    Data from \cite{winowiecki_thesis} is plotted against simulated results for (a), (d)-(f).
    }
    \label{fig:hMetal_Ni100_dt25920_ens_3x3}
\end{figure}

\begin{figure}[h]%
    \centering
    \includegraphics[width=1\textwidth]{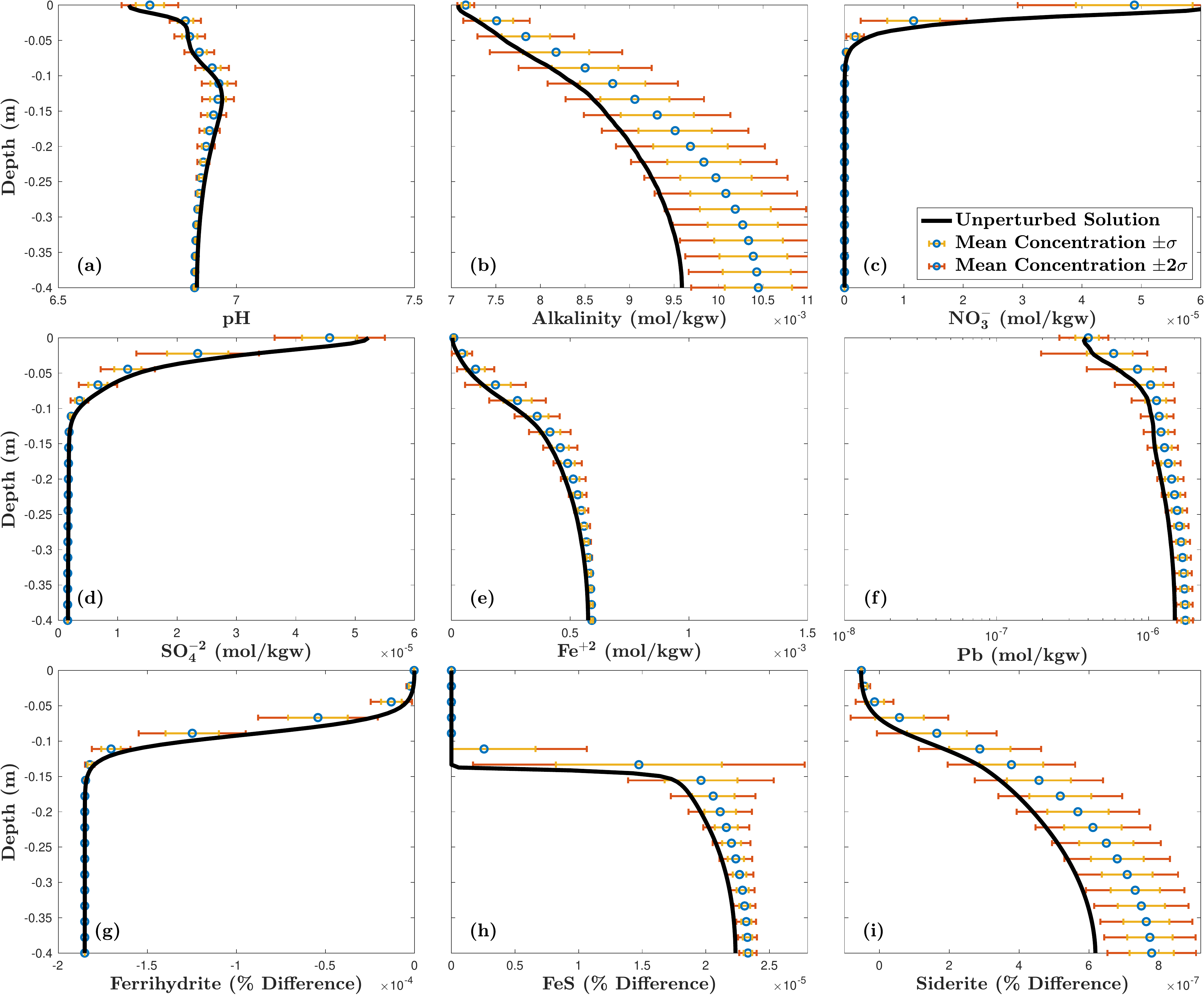}
    \caption{Ensemble results (100 realizations) for 100 randomly-spaced immobile particles and $\dt = 25920$.
    Concentrations are grouped into 19 bins, and mean values and error bars of $\pm 1$ and $\pm 2$ standard deviations are depicted for each bin.
    Final time concentrations ($t = 5$ years) of key chemical markers and pH are shown in plots (a)-(f) and percentage difference between initial and final amounts of secondary iron species is shown in plots (g)-(i).
    }
    \label{fig:hMetal_Ni100_dt25920_eBar_3x3}
\end{figure}

In this section, we consider the effect of spatial perturbations to the positions of immobile particles.
To do this, we set the positions of immobile particles according to random draws from a uniform distribution $\cU(0.0, -0.4)$ and set the initial concentrations on each immobile particle to be the same as in the previous, equally-spaced, case (see Table \ref{tab:hMetal_ICBC}).
An ensemble of 100 realizations is then conducted, each with a different spatial distribution of immobile particles.
As in Section \ref{ssub:IC_perturbations}, we perform simulations for $(N_I, \dt) = (100, 25920)$ and $(N_I, \dt) = (400, \{2592, 25920\})$, and find the combination $(N_I, \dt) = (100, 25920)$ to be sufficiently representative, so that we only depict and discuss those results.

In Figure \ref{fig:hMetal_Ni100_dt25920_ens_3x3}, we plot a selection of 3 realizations out of the 100 conducted, and we specifically choose these three realizations so as to demonstrate the spatial variability that may be induced by this type of perturbation.
Figure \ref{fig:hMetal_Ni100_dt25920_eBar_3x3} depicts the ensemble mean (blue markers) and $\pm 1$ and $\pm 2$ standard deviations of the concentration ensemble results (yellow and orange error bars, respectively).
In all plots, we compare to the unperturbed solution of Section \ref{sub:particle_tracking_results_base} (solid black line), and we also show the data from \cite{winowiecki_thesis} in selected plots.
As mentioned in the previous section, we bin the ensemble data into 19 equally-spaced bins to compute the statistics plotted in Figure \ref{fig:hMetal_Ni100_dt25920_eBar_3x3}.

\section{Discussion} 
\label{sec:discussion}

Here, we discuss the results of Section \ref{sec:results_Hmetal}.
Section \ref{sub:numerical_discussion} contains a numerical perspective on the results generated by the two perturbation methods presented in \ref{sub:perturbation_analysis}.
For ease of discussion, we will refer to the perturbations of immobile particle initial concentrations of Section \ref{ssub:IC_perturbations} as the CP (concentration perturbation) case and the perturbations of immobile particle spatial positions of Section \ref{ssub:randomly_spaced_immobile_particles} as the SP (spatial perturbation) case.
In contrast, Section \ref{sub:geochemical_discussion} is a geochemical discussion of the results.

\subsection{Numerical discussion of perturbation analysis} 
\label{sub:numerical_discussion}

One of the first things we notice, if we observe Figure \ref{fig:hMetal_Ni100_dt25920_rand_concs_ens_3x3}, is that the spread of the simulated results for the CP results is relatively narrow.
This is due to the fact that, while the perturbations discussed in this section do create an initially noisy distribution of reactants with areas of reactant scarcity, the overall mixing present in the system is unaltered.
Thus, the speed of reactions is not substantially reduced, and final time concentrations are perturbed symmetrically about the mean value of the unperturbed case, for the most part.
This behavior is more clearly evident in Figure \ref{fig:hMetal_Ni100_dt25920_rand_concs_eBar_3x3} where we see very close match between the ensemble means and the unperturbed case.
The exception to this, present in nearly every plot, is that the position of reaction fronts (spatial gradients) tends to be pushed upward, or delayed, as compared to the unperturbed reference solution.
This indicates that the initial concentration perturbations do have a measurable effect on mixing and reaction rates, though not as significant as in the SP case.

Another behavior of interest observed in Figure \ref{fig:hMetal_Ni100_dt25920_rand_concs_ens_3x3}, is that the concentration curves for the different realizations are, for the most part, parallel both to one another and to the reference unperturbed solution.
The exceptions to this generalization are Pb and FeS; Pb displays some minor crossing of curves, while FeS displays highly oscillatory behavior.
In the case of FeS, the reason for this is most likely related to the extremely low initial concentration of aqueous sulfur (in the unperturbed case, $C_0^{\text{Sulfur}} = \cO(10^{-28})$), and, as a result, any perturbation of this quantity will magnify the amount of FeS that precipitates.
As for Pb, we first note that these fluctuations are reflected in the data shown in Figure \ref{fig:arora_fig3_and_4}(f), leading to the conclusion that this type of perturbation may be more representative of the true conditions than the domain-wide constant initial condition employed in the unperturbed case.
Our explanation for the observed behavior in Figure \ref{fig:hMetal_Ni100_dt25920_rand_concs_ens_3x3}(f) is related to the ``shape'' of the concentration curve at the final time.
It is the only species that does not display a smooth reaction front that spans the domain and instead contains inflection points (other than pH, but the behavior of pH is much more smooth and tightly constrained).
As such, slight shifts in the position of the local maxima and minima of Pb (these can also be thought of as local reaction fronts) can have magnified downstream effects.

In the results displayed by Figure \ref{fig:hMetal_Ni100_dt25920_eBar_3x3}, we see a much greater degree of intra-ensemble concentration variability than that of Figure \ref{fig:hMetal_Ni100_dt25920_rand_concs_eBar_3x3}.
This indicates that a perturbed spatial distribution of solid species has a magnified impact on the results of HMLS reactive transport simulations, as compared to merely perturbing the initial concentrations.
Figure \ref{fig:hMetal_Ni100_dt25920_eBar_3x3} shows that alkalinity, Pb, and Siderite are quite sensitive to these spatial perturbations, and all display a high degree of variability throughout the domain.
Physically, this is due to several biotic reactions that consume acetate and an electron donor (e.g., O$_2$, NO$_3^{-}$, Fe$^{+3}$, and SO$_4^{-2}$) to produce bicarbonate, thereby increasing alkalinity.
For this reason, the primary electron donors, nitrate (NO$_3^-$) and sulfate $\left(\text{SO}_4^{2-}\right)$, display a similar degree of variability near the upper boundary $(x = 0)$, and Fe$^{+2}$ and Pb show a lesser degree of variability, also mainly focused near the upper boundary.
All species display the greatest variability in concentration where large magnitude gradients, or fronts, exist, likely because these fronts indicate regions in which a given species is out of chemical equilibrium after diffusing some small distance.
A sufficiently large gap between immobile particles will slow down reaction rates in that area because aqueous (mobile) species will take time to traverse this distance before they are ``eligible'' for reaction.
Such reasoning also explains the greater variability in alkalinity, as compared to other quantities because it displays a front that spans the entire domain at final time, while nitrate, sulfate, Fe$^{+2}$, and Pb are nearly constant throughout much of the domain.

One of the more important aspects of these SP simulations may be observed most clearly in Figure \ref{fig:hMetal_Ni100_dt25920_eBar_3x3}.
In particular, for species that display relatively stable final time behavior (e.g., pH, nitrate, sulfate, or ferrihydrite), their ensemble-mean value closely matches the unperturbed solution.
However, for species that display sharp gradients at final time, the ensemble mean tends to lag behind the unperturbed solution (in relation to the direction the species is traveling), confirming that spatial gaps slow down reaction fronts.

Additionally, in Figure \ref{fig:hMetal_Ni100_dt25920_ens_3x3}, we see that the depicted concentration curves are non-smooth, and do not parallel the unperturbed solution, as they do in Figure \ref{fig:hMetal_Ni100_dt25920_rand_concs_ens_3x3}.
This more closely approximates the noise and variability we see in the data, which is also non-smooth in space, within a single sample.
In fact, if we compare Figures \ref{fig:hMetal_Ni100_dt25920_rand_concs_ens_3x3}(f) and \ref{fig:hMetal_Ni100_dt25920_ens_3x3}(f), we see that the results in Figure \ref{fig:hMetal_Ni100_dt25920_ens_3x3}(f) more closely capture the oscillation and spread of the data.
As such, it is likely that the true distribution of solids involves a spatially-perturbed condition that alters mixing, as in the SP case, rather than just perturbed concentrations, as in the CP case.


\subsection{Geochemical discussion} 
\label{sub:geochemical_discussion}

Even in the relatively simple geochemical system explored here, heterogeneity can play a role in spatial distribution of geochemical processes.
Pb exhibits the greatest variation in aqueous concentrations within the original data, likely due to the sorption of Pb to mineral and organic surfaces controlling the concentrations.
In the original modeling of \emph{Arora et al.} and the modeling performed here, ferrihydrite is the only mineral surface Pb is allowed to sorb to, and ferrihydrite has a homogeneous surface area.
In natural systems, surface area of ferrihydrite can vary, and other mineral or organic surfaces may be available for metal sorption. Adding additional heterogeneity in these parameters may permit better fitting of the variations in concentrations, but the model fits the general distributions with higher concentrations in the subsurface, decreasing upward toward the water-sediment interface.
For all the parameters evaluated, the biggest differences between unperturbed and perturbed simulations are observed around the active reaction front, where consumption of oxygen through acetate oxidation changes redox conditions and drives reduction of aqueous Fe$^{+3}$ and subsequent dissolution of ferrihydrite.
The cascading effect of heterogeneity on myriad geochemical processes in a single system is observed in this case study as well.
When comparing simulation results from the unperturbed case and the immobile particle position perturbation case (SP), alkalinity shows the largest deviation with higher alkalinity concentrations at depth $> 0.2$ m in the perturbed case compared to the unperturbed case (Figure \ref{fig:hMetal_Ni100_dt25920_ens_3x3}(b)).
The higher concentrations of carbonate ions at depth lead to more siderite precipitation (Figure \ref{fig:hMetal_Ni100_dt25920_ens_3x3}(i)).
Carbonate is produced from the oxidation of acetate, suggesting that more acetate is oxidized and more oxygen is consumed when particles are distributed heterogeneously (randomly) than when they are distributed homogeneously (evenly).
As more oxygen is consumed, more iron is reduced and more ferrihydrite dissolves, releasing Pb into solution and more sulfate is reduced to HS$^-$ and leads to FeS precipitation (Figure \ref{fig:hMetal_Ni100_dt25920_ens_3x3}(h)).
Heterogeneous distribution of solid phases also impacts the positioning of the reaction front--in this case with shallower reaction fronts where minerals are heterogeneously distributed.
These results highlight the importance of considering heterogeneity in geochemical systems when the spatial distributions of geochemical processes are important \cite[e.g.,][]{heewon_alexis_scale_effect,heewon_alexis_hetero_control}.



\section{Conclusions} 
\label{sec:conclusions_hMetal}

In this work, we have applied the miRPT algorithm of \cite{Schmidt_fluid_solid} to model a benchmark reactive transport problem involving heavy metal cycling in lake sediments (the HMLS system).
This system was modeled using Eulerian methods in \cite{Arora2015}, and those authors achieved favorable and nearly identical results for all considered methods.
However, the smooth curves produced by Eulerian models fail to capture the variability inherent to the data (see Figure \ref{fig:arora_fig3_and_4}).
In the unperturbed base case implementation (Section \ref{sub:particle_tracking_results_base}), we first recreated the results of a corresponding Eulerian model with very close match.
The main differences are a varying resolution of sharp gradients and differing boundary concentrations of nitrate and sulfate.
The former is typical of varying levels of discretization, and the latter is due to the influence of time step length on the ability of these species to reach equilibrium.
As such, we conclude that the miRPT model, in the unperturbed case, is capable of capturing the same behavior as a corresponding Eulerian model.

The primary focus herein was to investigate the impact of imperfectly-mixed reactants on the behavior of a complicated geochemical system.
Eulerian models are not as well suited to represent this kind of physical heterogeneity because they can only do so by increasing discretization, leading to a more restrictive time step and a greater number of expensive chemistry calculations during each time step.
In contrast, the Lagrangian model used in this work is uniquely suited to this task because the number of chemistry calculations per time step can be fixed by selecting the number of immobile particles within a simulation while still increasing spatial resolution of aqueous species, practically for free, by increasing the number of mobile particles that appear.
To explore the effects of imperfect mixing, we perturbed the representation of solid species in our model using two different approaches.
To demonstrate the typical method for perturbing a reactive transport simulation, in Section \ref{ssub:IC_perturbations} we randomly varied the initial concentrations of reactants on evenly-spaced immobile particles.
This introduced a noisy initial condition to the problem and can also be achieved using Eulerian methods.
The results of this perturbed experiment failed to capture the variability and noise in the data (see Figure \ref{fig:hMetal_Ni100_dt25920_rand_concs_ens_3x3}), as the variation in results was relatively minor, and most results smoothly parallel the unperturbed solution, which is closely captured by the ensemble mean.
This is because the concentration perturbations initially effect the state of the system but are quickly mitigated over the course of the simulation, and overall mixing in the system is not affected for large times.

The more interesting result is found in Section \ref{ssub:randomly_spaced_immobile_particles}, wherein we spatially perturbed simulations by randomly varying the positions of immobile particles.
This is a method that cannot be explored using Eulerian methods because neighboring grid points communicate during every time step, which makes it impossible to create the persistent zones of poor mixing that are induced by spatial gaps between immobile particles.
To be clear, Eulerian methods can certainly employ a stochastically-perturbed grid to mimic part of this behavior; however, without the mobile particles that carry the reactants through space, they will not fully capture this poor mixing due to reactant segregation.
Additionally, introducing the perturbations (of either sort) into the particle-tracking simulations required no change to the algorithm, while in the case of Eulerian models, this would be a signifcantly more complicated endeavor than employing an equally-spaced grid.
Finally, in order to ensure the stability of a spatially-perturbed Eulerian simulation, one would need to choose a time step length corresponding to the smallest spacing in each grid realization, leading to unpredictable and possibly prohibitive run times.

The results of the spatially-perturbed experiment are compelling because we see much wider variation in the intra-ensemble results (see Figure \ref{fig:hMetal_Ni100_dt25920_ens_3x3}), and this variance is an important feature of the data that neither the single-solution Eulerian models, nor the alternative perturbation method can fully capture.
For this reason, the spatial perturbation method captures an important real-world feature that is neglected by other modeling and simulation methods.
Specifically, we are able to capture the slowdown in reaction speed that is induced by poorly-mixed conditions.
Ultimately, we conjecture, though it remains to be rigorously proven, that the perturbations in immobile particle position correspond mathematically to perturbations in the magnitude of diffusion present in the system, a reasonable and necessary modeling choice when trying to capture the irregular behaviors induced by the inhomogeneity of porous media.


\section{Acknowledgments} 
\label{sec:acknowledgments}

The authors thank Arora et al., the authors of \cite{Arora2015}, for sharing their PHREEQC input and database files, which were indispensable to this work.

This work was supported by the US Army Research Office under contract/grant number W911NF-18-1-0338; the National Science Foundation under awards EAR-1417145 and DMS-1614586; and the DOE Office of Science under award DE-SC0019123.


\bibliography{hMetals_bibl}

\end{document}